\documentclass[preprint,12pt]{elsarticle}%
\usepackage{amssymb}
\usepackage{amsmath}
\usepackage{amsfonts}
\usepackage{graphicx}
\usepackage{epsfig}%
\providecommand{\U}[1]{\protect\rule{.1in}{.1in}}
\journal{Chemical Physics Letters}
\begin{document}
\begin{frontmatter}
\title{Monte Carlo evaluation of the equilibrium isotope effects using the
Takahashi-Imada factorization of the Feynman path integral}
\author{Marcin Buchowiecki}
\address{Institute of Physics, University of Szczecin, Wielkopolska 15\\
Szczecin, 70-451, Poland\\
mabuch@wmf.univ.szczecin.pl}
\author{Ji\v{r}\'{\i} Van{\'{\i}}{\v{c}}ek}
\address{Laboratory of Theoretical Physical Chemistry, Institut des Sciences et
Ing\'{e}nierie Chimiques, \'{E}cole Polytechnique F\'{e}d\'{e}rale de Lausanne
(EPFL), CH-1015 Lausanne, Switzerland\\
jiri.vanicek@epfl.ch}
\begin{abstract}
The Feynman path integral approach for computing equilibrium isotope effects
and isotope fractionation corrects the approximations made in standard
methods, although at significantly increased computational cost. We describe
an accelerated path integral approach based on three ingredients: the
fourth-order Takahashi-Imada factorization of the path integral, thermodynamic
integration with respect to mass, and centroid virial estimators for relevant
free energy derivatives. While the first ingredient speeds up convergence to
the quantum limit, the second and third improve statistical convergence. The
combined method is applied to compute the equilibrium constants for isotope
exchange reactions H$_{2}+\,$D$\,\rightleftharpoons\,$H\thinspace+ HD and
H$_{2}+\,$D$_{2}\,\rightleftharpoons\,2$HD.
\end{abstract}
\end{frontmatter}


\section{\label{sec:intro}Introduction}

The equilibrium isotope effect (EIE), defined as the effect of isotopic
substitution on chemical equilibrium, is amongst the most important tools for
studying reaction mechanisms and equilibria
\cite{Wolfsberg_VanHook_Paneth:2010}. In certain situations, e.g., in
composite mechanisms with a pre-equilibrium, the EIE can be the main
determinant of the kinetic isotope effect. Although EIE research has a long
history, until recently most theoretical calculations were based on the
assumption of separability of rotations and vibrations, on the rigid rotor
approximation for rotations, and on the harmonic treatment of vibrations
\cite{Wolfsberg_VanHook_Paneth:2010}. More rigorous EIE treatments avoiding
these three approximations are recent
\cite{Lynch:2005,Zimmermann_Vanicek:2009,Zimmermann_Vanicek:2010b,Azuri:2011,Perez_Lilienfeld:2011,Ceriotti_Manolopoulos:2012,Ceriotti_Markland:2013}
and based on the Feynman path integral
\cite{feynman:1965,kleinert:2004,ceperley:1995} representation of the
partition function.

One of these rigorous approaches \cite{Zimmermann_Vanicek:2009}, the extension
of which is the main goal of this article, evaluates the required\ ratio of
partition functions via thermodynamic integration with respect to the isotope
mass (used originally \cite{Yamamoto_Miller:2005,vanicek:2005,vanicek:2007} in
the context of quantum instanton \cite{miller:2003} calculation of kinetic
isotope effects) and via Monte Carlo sampling of the discretized path
integral. The method of Ref. \cite{Zimmermann_Vanicek:2009} was successfully
used with ab initio potentials for calculating the EIE in $[1,5]$ sigmatropic
hydrogen shift reactions
\cite{Zimmermann_Vanicek:2009,Zimmermann_Vanicek:2010b}. The computational
cost was drastically lowered by combining the harmonic estimate of the EIE
based on accurate ab initio electronic structure calculations together with
path integral Monte Carlo simulations based on force fields or semiempirical
potentials for calculating the anharmonicity and other quantum corrections. In
both Refs. \cite{Zimmermann_Vanicek:2009,Zimmermann_Vanicek:2010b}, only the
standard primitive approximation for the path integral was used.


Lynch, Mielke, and Truhlar \cite{Lynch:2005} evaluated accurate equilibrium
isotope effects on H$_{2}$O$_{2}$ as ratios of partition functions computed
with the enhanced same-path extrapolation variant \cite{Mielke_Truhlar:2003}
of the trapezoidal Trotter Fourier path integral Monte Carlo method
\cite{Coalson:1986}. Another method, proposed by Major and coworkers
\cite{Azuri:2011} combines the quantized classical path approach of Hwang and
Warshel \cite{Hwang_Warshel:1993} with a perturbation expression using the
staging algorithm. After comparing implementations with the primitive,
Takahashi-Imada \cite{Takahashi_Imada:1984}, and Chin \cite{Chin:2004}
factorizations of the discretized path integral, the authors applied their
method to compute the EIE on the keto-enol tautomerism in alanine racemase
\cite{Azuri:2011}. Perez and von Lilienfeld generalized the calculation of
EIEs using \textquotedblleft alchemical\textquotedblright\ transformations\ in
which the isotope masses are changed simultaneously with the interaction
potential \cite{Perez_Lilienfeld:2011}. Ceriotti and Manolopoulos sped up
EIE\ calculations by combining path integral molecular dynamics with a
generalized Langevin equation \cite{Ceriotti_Manolopoulos:2012}. Very
recently, Ceriotti and Markland accelerated the thermodynamic integration with
respect to mass by \textquotedblleft flattening\textquotedblright\ the
integral with a nonlinear change of variables \cite{Ceriotti_Markland:2013}.


In general, methods based on path integrals are accurate, but also very
computationally expensive when compared with the less rigorous approaches.
Consequently, any extension that significantly speeds up the convergence of
path integral\ simulations without compromising accuracy is highly desirable.
The aim of this paper is to accelerate the convergence of the Ref.
\cite{Zimmermann_Vanicek:2009} method to the quantum limit through combination
with the fourth-order Takahashi-Imada factorization
\cite{Takahashi_Imada:1984} of the path integral. The smaller error in
Takahashi-Imada factorization allows discretization of the path integral
with\ fewer imaginary time slices than required by the primitive approximation
used in Ref. \cite{Zimmermann_Vanicek:2009}. In addition to lowering the
systematic error by the Takahashi-Imada\ factorization, we also decrease the
statistical error by implementing the centroid virial estimator for the free
energy derivative with respect to mass and specific for the Takahashi-Imada
factorization. The improved methodology is tested on two isotope exchange
reactions on the BKMP2 potential energy surface \cite{Boothroyd:1996}. In
contrast to Ref. \cite{Zimmermann_Vanicek:2009}, where the discretized path
integral was sampled with the path integral molecular dynamics, here the
sampling is performed with the path integral Monte Carlo method.

The remainder of the article is organized as follows: Section~\ref{sec:theory}
explains how to evaluate EIEs via the thermodynamic integration, describes the
implementation of this method using the Takahashi-Imada\ factorization of the
path integral, and presents the derivation of relevant thermodynamic and
centroid virial estimators. The methodology described in Sec.~\ref{sec:theory}
is applied to two different chemical reactions in Sec.~\ref{sec:results},
while Sec.~\ref{sec:conclusions} concludes the paper.

\section{\label{sec:theory}Theory}

\subsection{\label{sec:theory_Th-Int}Thermodynamic integration with respect to
mass}

The EIE is defined as the ratio EIE $=K_{l}/K_{h},$where $K_{l}$ and $K_{h}$
are the equilibrium constants of two chemical reactions that involve lighter
($l$) and heavier ($h$) isotopologs, but are otherwise identical. In terms of
the molecular partition functions for the products ($P$) and reactants ($R$),
the EIE is expressed as \cite{Mayer_Mayer:1977}%
\begin{equation}
\text{EIE}=\frac{Q_{l}^{P}/Q_{l}^{R}}{Q_{h}^{P}/Q_{h}^{R}}. \label{eie}%
\end{equation}
A more convenient form of the last expression, namely,
\begin{equation}
\text{EIE}=\frac{Q_{h}^{R}/Q_{l}^{R}}{Q_{h}^{P}/Q_{l}^{P}}, \label{eie_q}%
\end{equation}
suggests that an evaluation via the thermodynamic integration
\cite{frenkel:2002} with respect to the isotopic mass is possible
\cite{Zimmermann_Vanicek:2009,vanicek:2005}. In this method, the ratios of
partition functions in Eq.~(\ref{eie_q}) are computed from the reduced free
energy difference $\Delta F$ as
\[
\frac{Q_{h}}{Q_{l}}=\frac{s_{l}}{s_{h}}\exp\left(  -\beta\Delta F\right)  ,
\]
where $s_{l}$ and $s_{h}$ denote the symmetry numbers of the two isotopologs
and $\beta=1/(k_{B}T)$ is the inverse temperature. By factoring out the
symmetry numbers, we take into account the exchange symmetry of the
indistinguishable hydrogen nuclei approximately and $\Delta F$ can be computed
as for distinguishable particles. The symmetry numbers reflect the number of
indistinguishable orientations of the molecule in which hydrogen atoms are
exchanged by rotation
\cite{Wolfsberg_VanHook_Paneth:2010,Mayer_Mayer:1977,Wolfsberg:1972}. The
value of EIE stemming from the symmetry numbers alone is referred to as a
purely statistical isotope effect, which is equal to the high temperature
approximation described below. The treatment of indistinguishability by
symmetry factors is valid even at temperatures much lower than those
considered here \cite{feynman:1965,ceperley:1995}. (To go beyond this
approximation is possible, but requires Feynman paths connecting different
particles \cite{feynman:1965,ceperley:1995}; the contributions of such paths
to the Feynman path integral are strongly suppressed except at temperatures
close to absolute zero.)

The reduced free energy difference $\Delta F$ is evaluated as the integral
\begin{equation}
\Delta F=\int_{0}^{1}d\lambda\frac{dF(\lambda)}{d\lambda} \label{q_r}%
\end{equation}
of the derivative of $F$ with respect to a parameter $\lambda$ interpolating
between the masses of isotopes in the two isotopologs. For convenience, we
choose this interpolation to be linear:%
\begin{equation}
m_{i}(\lambda)=(1-\lambda)m_{l,i}+\lambda m_{h,i}. \label{mass_interpol}%
\end{equation}

Finally, note that for an isotopomerization reaction (a reaction in which $R$
and $P$ are isotopomers, i.e., involve the same numbers of all isotopes), the
EIE can be simply defined as the equilibrium constant: EIE $=K=Q^{P}/Q^{R}.$


\subsection{\label{sec:theory_TI-PIMC}Monte Carlo evaluation of the free
energy derivatives via the Takahashi-Imada factorization of the path integral}

The path integral formalism \cite{feynman:1965,kleinert:2004,ceperley:1995}
allows for a rigorous representation of the quantum partition function
$Q(\beta)=\operatorname{Tr}[\exp{(-\beta\hat{H})}]$ without knowledge of the
eigenstates of the Hamiltonian operator. Using the Lie-Trotter theorem
\cite{Trotter:1959}, the trace on the right side of this expression can be
factorized as%
\begin{equation}
\operatorname{Tr}(e^{{-\beta\hat{H}}})=\operatorname{Tr}\{[e^{{-(\beta
/P)\hat{T}}}e^{{-(\beta/P)\hat{V}}}]^{P}\}+\mathcal{O}(\beta^{3}P^{-2}),
\label{pa}%
\end{equation}
where $\hat{T}$ and $\hat{V}$ are the kinetic and potential energy operators,
and $P$ is the Trotter number (or number of imaginary time slices or
\textquotedblleft beads\textquotedblright). Setting $P=1$ in Eq.~(\ref{pa})
yields the classical partition function, whereas the limit $P\rightarrow
\infty$ gives the exact quantum partition function. Factorization (\ref{pa})
is usually called the \emph{primitive approximation} (PA). Due to its
simplicity, it is the most widely used path-integral factorization and was
used for the calculation of the EIEs in Ref.~\cite{Zimmermann_Vanicek:2009}.

Our main goal is to demonstrate that the convergence of the EIE\ calculation
with respect to $P$ is significantly accelerated by employing the
\emph{Takahashi-Imada} (TI) factorization \cite{Takahashi_Imada:1984},
\begin{equation}
\operatorname{Tr}(e^{{-\beta\hat{H}}})=\operatorname{Tr}\{[e^{{-(\beta
/P)\hat{T}}}e^{{-(\beta/P)\hat{V}}_{\text{eff}}}]^{P}\}+\mathcal{O}\left(
\beta^{5}P^{-4}\right)  , \label{tia}%
\end{equation}
where $\hat{V}_{\text{eff}}$ is an effective potential energy operator
obtained by augmenting the true potential energy $\hat{V}$ with a TI term
$\hat{V}_{\text{TI}},$
\begin{align}
\hat{V}_{\text{eff}}  &  =\hat{V}+\hat{V}_{\text{TI}},\label{veff}\\
\hat{V}_{\text{TI}}  &  =\frac{1}{24}\left(  \frac{\beta}{P}\right)  ^{2}%
[\hat{V},[\hat{T},\hat{V}]]. \label{vti}%
\end{align}
In the coordinate representation,
\[
V_{\text{TI}}(\mathbf{r})=\frac{1}{24}\hbar^{2}\left(  \frac{\beta}{P}\right)
^{2}\sum_{i=1}^{N}\frac{1}{m_{i}}\left(  \frac{\partial V}{\partial
\mathbf{r}_{i}}\right)  ^{2}.
\]
\textbf{where }$r=(r_{1},...,r_{N})$\textbf{ is a collective notation for the
coordinates of all }$N$\textbf{ atoms.} Having accuracy of higher order in
$P$, the TI factorization is more accurate than the PA for a given
(sufficiently large) $P.$ As a consequence, the TI factorization achieves a
desired accuracy with a smaller Trotter number $P$ than the PA. This is an
advantage, particularly at low temperatures, where the Trotter number required
for simulation convergence with the PA\ can be very large
\cite{Azuri:2011,Takahashi_Imada:1984,Jang:2001,Brualla:2004,Perez_Tuckerman:2011,Hellmann:2011}%
.

Both PA (\ref{pa}) and TI factorization (\ref{tia}) are expressed exactly in
the coordinate representation. In the TI scheme,
\begin{equation}
Q_{\text{TI,}P}(N,V,T)=C\int d\{\mathbf{r}^{(s)}\}\,\exp[-\beta\Phi
_{\text{eff}}(\{\mathbf{r}^{(s)}\})], \label{pi_discr}%
\end{equation}
where $\{\mathbf{r}^{(s)}\}$ stands for $(\mathbf{r}^{(1)},\ldots
,\mathbf{r}^{(P)})$; the prefactor $C$ and effective \textquotedblleft polymer
chain\textquotedblright\ potential $\Phi_{\text{eff}}$ are
\begin{align}
C  &  =\left(  \frac{P}{2\pi\hbar^{2}\beta}\right)  ^{3NP/2}\prod_{i=1}%
^{N}m_{i}^{3P/2}\text{ and}\nonumber\\
\Phi_{\text{eff}}(\{\mathbf{r}^{(s)}\})  &  =\frac{P}{2\hbar^{2}\beta^{2}}%
\sum_{i=1}^{N}m_{i}\sum_{s=1}^{P}(\mathbf{r}_{i}^{(s)}-\mathbf{r}_{i}%
^{(s+1)})^{2}\label{Phi_eff}\\
&  +\frac{1}{P}\sum_{s=1}^{P}V_{\text{eff}}(\mathbf{r}^{(s)}).\nonumber
\end{align}
The discretized path integral~(\ref{pi_discr}) is evaluated with either the
path integral Monte Carlo (PIMC) or path integral molecular dynamics, with
both approaches having advantages and disadvantages. While the path integral
molecular dynamics appears to be the method of choice for \emph{ab initio}
path integral calculations \cite{Perez_Tuckerman:2011}, its PA implementation
requires force and energy evaluations, whereas the PIMC needs only energies.
The TI factorization implemented in a PIMC code requires only energies and
forces, whereas for path integral molecular dynamics energies, forces,
\emph{and} second derivatives of the energies must be calculated. \textbf{The
second derivative requirement was circumvented by Jang and coworkers
\cite{Jang:2001}, Yamamoto \cite{Yamamoto:2005}, and Perez and Tuckerman
\cite{Perez_Tuckerman:2011} with an elegant trick employing the PA as a
reference potential; however, this assumes that the TI correction (\ref{vti})
is small. Ceriotti and coworkers \cite{Ceriotti:2012} showed recently that
this re-weighting procedure fails in systems with many degrees of freedom due
to the growth of the statistical error of the re-weighted average.}

Therefore we focus on the PIMC implementation of the TI\ factorization: In
addition to not requiring second derivatives of the potential energy, its
implementation into existing PIMC codes is straightforward, the main change
being that both the random walk and the estimators use $\Phi_{\text{eff}}$
instead of the $\Phi$ originally employed in the PA. Below we list only the
new, TI expressions; the corresponding PA\ expressions can be obtained by
setting $V_{\text{TI}}=0$. Moreover, they have already been derived in
Ref.~\cite{vanicek:2007}.

PIMC simulations \cite{ceperley:1995} are based on the sampling of the full
$3NP$-dimensional configuration space of the discretized path-integral
representation (\ref{pi_discr}) of the partition function. The
polymer-chain\ potential energy $\Phi_{\text{eff}}$ from Eq.~(\ref{Phi_eff})
determines the sampling weight $W=\exp(-\beta\Phi_{\text{eff}})$. In compact
notation, the thermodynamic average $\left\langle A\right\rangle _{T}$ of an
operator $\hat{A}$ at temperature $T$ is obtained as a Monte Carlo average
$\left\langle A\right\rangle _{T}\approx\langle A_{\text{E}}(\{\mathbf{r}%
^{(s)}\})\rangle_{W}$, where $A_{\text{E}}$ is an estimator for $\hat{A}$ and
$W$ is the sampling weight. The path-integral estimator for the derivative of
the free energy in Eq.~(\ref{q_r}) is obtained most easily by substituting the
path-integral representation (\ref{pi_discr}) into the equation $-\beta
dF/d\lambda=d\log Q(\lambda)/d\lambda$. This results in the thermodynamic
estimator,%
\[
-\beta\frac{dF_{\text{TI,TE}}(\lambda)}{d\lambda}=\sum_{i=1}^{N}\frac{dm_{i}%
}{d\lambda}\left(  \frac{3P}{2m_{i}}-\beta\frac{d\Phi_{\text{eff}}}{dm_{i}%
}\right)  ,
\]
where%
\[
\beta\frac{d\Phi_{\text{eff}}}{dm_{i}}=\sum_{s=1}^{P}\left[  \frac{P}%
{2\hbar^{2}\beta}\left(  \mathbf{r}_{i}^{(s)}-\mathbf{r}_{i}^{(s+1)}\right)
^{2}+\frac{\beta}{P}\frac{dV_{\text{TI}}(\mathbf{r}^{(s)})}{dm_{i}}\right]  .
\]
The problem with this estimator is the growth of its statistical error with
$P$, preventing convergence to the quantum limit. A similar effect was
observed for the thermodynamic estimator in the PA and was fixed by using the
centroid virial estimator for $dF_{\text{PA}}(\lambda)/d\lambda$
\cite{vanicek:2007}, which is a generalization of the centroid virial
estimator for the kinetic energy \cite{Herman:1982,parrinello:1984}. As in the
PA used in Ref.~\cite{vanicek:2007}, in the TI scheme this estimator can be
derived by mass-scaling the coordinates \cite{Predescu2004} and subtracting
the centroid coordinate $\mathbf{r}^{C}:=P^{-1}\sum_{s=1}^{P}\mathbf{r}^{(s)}$
in Eq.~(\ref{pi_discr}). This shifts the dependence on mass completely from
the kinetic to the potential energy and the application of the equation
$-\beta dF/d\lambda=d\log Q(\lambda)/d\lambda$ yields the desired TI centroid
virial estimator:
\begin{align}
-\beta\frac{dF_{\text{TI,CVE}}(\lambda)}{d\lambda}  &  =\frac{D}{2}\sum
_{i=1}^{N}\frac{1}{m_{i}}\frac{dm_{i}(\lambda)}{d\lambda}\label{est}\\
&  -\frac{\beta}{P}\sum_{s=1}^{P}\left[ \frac{dV_{\text{eff}}(\{\mathbf{r}%
^{(s)}(\Delta\lambda)\})}{d\Delta\lambda}-\frac{\hbar^{2}}{24}\left(
\frac{\beta}{P}\right)  ^{2}\sum_{i=1}^{N}\frac{1}{m_{i}^{2}}\frac{dm_{i}%
}{d\lambda}\left(  \frac{\partial V}{\partial\mathbf{r}_{i}}(\mathbf{r}%
_{i}^{(s)})\right)  ^{2}\right] ,\nonumber\\
\mathbf{r}_{i}^{(s)}(\Delta\lambda):=  &  \mathbf{r}_{i}^{C}+\left[
\frac{m_{i}(\lambda)}{m_{i}(\lambda+\Delta\lambda)}\right]  ^{1/2}%
(\mathbf{r}_{i}^{(s)}-\mathbf{r}_{i}^{C}).\nonumber
\end{align}
It is essential to evaluate the $\Delta\lambda$ derivative in the last
expression by finite difference \cite{Predescu2004}. Evaluating this
derivative analytically would result in expressions involving the gradient of
$V_{\text{eff}}$ and requiring the second derivatives of $V$, which would
eliminate the advantage of the TI\ PIMC. \textbf{In general, the cost of
evaluating the estimators can be made negligible in comparison with the cost
of the random walk because, due to correlations between samples, it is not
necessary to sample at each Monte Carlo step.}


\subsection{Standard approach based on harmonic approximation}

In the numerical examples our results are compared with the standard approach
\cite{Wolfsberg_VanHook_Paneth:2010} based on the \textquotedblleft harmonic
approximation.\textquotedblright\ By harmonic approximation we mean an
ensemble of the following three approximations:\ separability of rotations and
vibrations, harmonic approximation for vibrations, and rigid rotor
approximation for rotations. An overall EIE can be decomposed into elementary
isotope effects (IEs), i.e., simple ratios of molecular partition functions:
IE $:=Q_{l}/Q_{h}$. Within the harmonic approximation, the Teller-Redlich
theorem yields \cite{Wolfsberg_VanHook_Paneth:2010}%
\begin{equation}
\text{IE}_{\text{HA}}=\text{IE}_{\text{HA, }T\rightarrow\infty}\prod
_{n=1}^{3N-6}\frac{x_{l,n}}{x_{h,n}}\frac{1-e^{-x_{h,n}}}{1-e^{x_{l,n}}%
}e^{-(x_{l,n}-x_{h,n})/2}, \label{ha}%
\end{equation}
where $n$ runs over the vibrational degrees of freedom, $x_{n}=\beta
\hbar\omega_{n}$, and $\omega_{n}$ is the angular frequency of the $n$th
vibration. The prefactor
\begin{equation}
\text{IE}_{\text{HA, }T\rightarrow\infty}=\frac{s_{h}}{s_{l}}\prod_{i=1}%
^{N}\left(  \frac{m_{l,i}}{m_{h,i}}\right)  ^{3/2} \label{ha_highT}%
\end{equation}
is the high temperature limit of IE$_{\text{HA}}$. The low temperature limit
of IE$_{\text{HA}}$ (\ref{ha})\ is
\begin{equation}
\text{IE}_{\text{HA, }T\rightarrow0}=\text{IE}_{\text{HA, }T\rightarrow\infty
}\prod_{n=1}^{3N-6}\frac{x_{l,n}}{x_{h,n}}e^{-(x_{l,n}-x_{h,n})/2}.
\label{ha_lowT}%
\end{equation}
The above expressions apply to nonlinear molecules, having $3N-6$ vibrational
degrees of freedom. Yet, these expressions remain valid in general when taking
into account that the number of vibrations is zero for atoms and $3N-5$ for
linear molecules.

\section{\label{sec:results}Numerical examples}

\subsection{Reactions}

Our methodology is tested numerically by evaluating the EIEs on two chemical
reactions, the first EIE\ being the equilibrium constant for the
isotopomerization reaction%
\begin{equation}
\text{H}_{2}+\text{D}\rightleftharpoons\text{H}+\text{HD,} \label{reaction1}%
\end{equation}
with the interesting part of the isotope effect being its deviation from the
purely statistical value of $2$. Since the mass interpolation
(\ref{mass_interpol}) allows changing masses of several atoms simultaneously,
EIE$_{1}$ can be calculated in a direct way as
\begin{equation}
\text{EIE}_{1}^{\text{direct}}=Q_{\text{HD}+\text{H}}/Q_{\text{H}_{2}%
+\text{D}}. \label{eie1-direct}%
\end{equation}
For this calculation the free energy estimator must take into account a
simultaneous transformation H$\,\dashrightarrow\,$D in H$_{2}$ and
D$\,\dashrightarrow\,$H in D. Sometimes, however, it is useful to separate the
contributions to the EIE\ from these two transformations and write:
\begin{equation}
\text{EIE}_{1}=\frac{Q_{\text{HD}+\text{H}}/Q_{\text{H}_{2}+\text{H}}%
}{Q_{\text{H}_{2}+\text{D}}/Q_{\text{H}_{2}+\text{H}}}, \label{eie1}%
\end{equation}
where both the numerator and denominator were divided by $Q_{\text{H}%
_{2}+\text{H}}$. Finally, it is possible to write EIE (\ref{eie1-direct}) as
EIE\thinspace$=(Q_{\text{HD}}/Q_{\text{H}_{2}})(Q_{\text{H}}/Q_{\text{D}})$,
where the first factor can be computed in a much simpler, 3-dimensional
simulation (with one vibrational and two rotational coordinates) and the
second factor evaluated analytically. We choose a more tedious, but more
general approach expressed in Eqs. (\ref{eie1-direct})\ or (\ref{eie1}) and
use nine Cartesian coordinates so that our methodology remains unchanged for
unimolecular reactions (which have no asymptotic region with noninteracting
reactants) and for large molecules (for which transformation to internal
coordinates does not simplify the calculation). Last, a nine-dimensional
calculation is a more stringent test of the methodology from
Sec.~\ref{sec:theory}.

The second EIE is an equilibrium constant for the isotopomerization reaction
\begin{equation}
\text{H}_{2}+\text{D}_{2}\rightleftharpoons2\text{HD}, \label{reaction2}%
\end{equation}
having a statistical value of $4$. Although this takes place on a different
potential energy surface, the EIE can be cast into a form suitable for the
BKMP2 potential energy surface:%
\begin{equation}
\text{EIE}_{2}=\frac{Q_{\text{HD}+\text{HD}}}{Q_{\text{H}_{2}+\text{D}_{2}}%
}=\frac{Q_{\text{HD}}\,Q_{\text{HD}}}{Q_{\text{H}_{2}}\,Q_{\text{D}_{2}}%
}=\frac{Q_{\text{HD}+\text{H}}/Q_{\text{D}_{2}+\text{H}}}{Q_{\text{H}%
_{2}+\text{H}}/Q_{\text{HD}+\text{H}}}. \label{eie2}%
\end{equation}
In the last step, both the numerator and denominator were multiplied by
$Q_{\text{H}}^{2}$. \textbf{The second and third equalities hold because, in
the low-pressure, low-concentration limit, the reactants and products are
noninteracting in the asymptotic regions.}

\subsection{Computational details}

All PIMC simulations were performed in Cartesian coordinates $(\mathbf{r}%
_{1},...,\mathbf{r}_{N})$. The random walk employed the staging algorithm
\cite{sprik:1985} to move half of the beads at once. The thermodynamic
integral of Eq.~(\ref{q_r}) was evaluated with the Simpson method. The Trotter
number was increased proportionally to the inverse temperature. Using the PA,
in particular, the Trotter number was increased from $P=30$ at $1000~$K to
$P=160$ at $200~$K. Approximately the same convergence was obtained in the TI
scheme when $P$ was increased from $P=8$ at $1000~$K to $P=48$ at $200~$K. The
simulations had an overall length of $10^{7}$ steps, of which $25\%$ were a
warm-up (i.e., equilibration of the system). Since the BKMP2 potential energy
surface provides both energies and forces, the forces required in the TI
scheme were computed analytically.

\textbf{Harmonic approximation results were based on the frequencies from the
BKMP2 PES, where the force constant was calculated as the numerical second
derivative.}


\subsection{Results}

\begin{figure}[ptbh]
\includegraphics[width=\hsize]{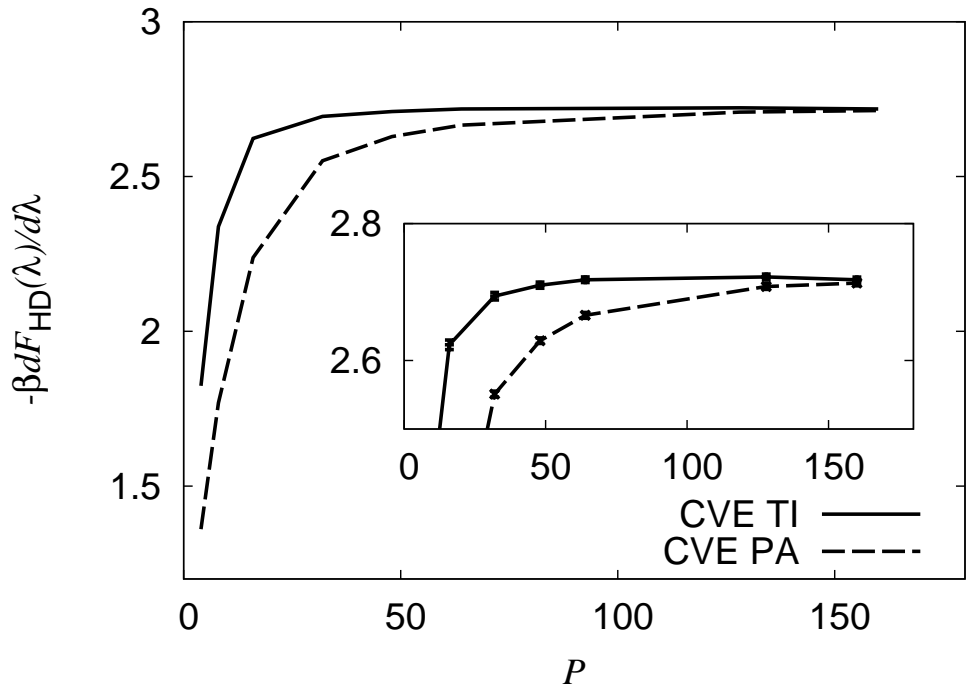}\caption{Free energy derivative as a
function of the Trotter number $P$. The free energy derivative corresponds to
the transformation H$_{2}+\,$H$\,\dashrightarrow\,$HD\,+\,H at $T=200~$K and
is evaluated with the centroid virial estimator (CVE) at $\lambda=0.5$. The
Takahashi-Imada (TI) factorization converges to the quantum limit much faster
than the primitive approximation (PA). The inset shows a detail including
statistical error bars. }%
\label{fig:dFdl}%
\end{figure}

Prior to calculating EIEs themselves, we analyzed convergence with respect to
the Trotter number $P$ of the derivative of the free energy $F_{\text{HD}}$
corresponding to the \textbf{isotopic \textquotedblleft
alchemical\textquotedblright\ transformation H}$_{2}+\,$\textbf{H\thinspace
}$\dashrightarrow\,$\textbf{HD\thinspace}$+\,$\textbf{H} in the numerator of
Eq.~(\ref{eie1}). Figure~\ref{fig:dFdl} demonstrates that both the PA and TI
schemes converge to the same results for $P\rightarrow\infty$, yet the TI
factorization converges to the quantum result for significantly smaller $P$
values, confirming the objective of our paper.

\begin{figure}[ptbh]
\includegraphics[width=\hsize]{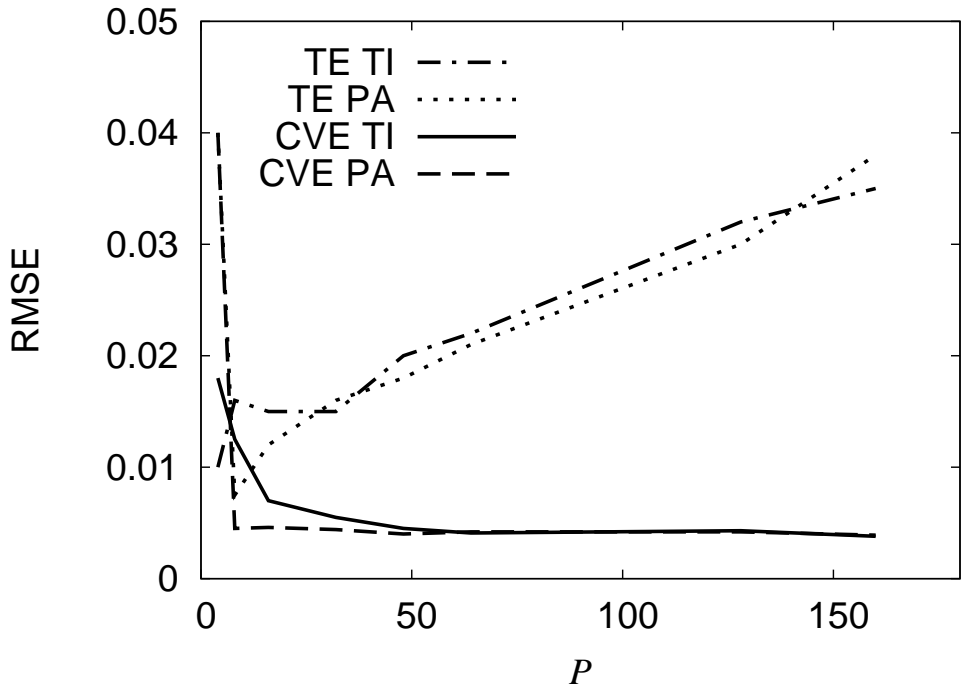}
\caption{Statistical root mean square errors (RMSE) of various estimators for
the free energy derivative $-\beta dF_{\text{HD}}(\lambda)/d\lambda$ from
Fig.~\ref{fig:dFdl} as a function of the Trotter number $P$. Both with the
primitive approximation (PA) and Takahashi-Imada (TI) factorization, the RMSE
is approximately independent of $P$ for the centroid-virial estimators (CVE),
whereas it grows with $P$ for the thermodynamic estimators (TE).}%
\label{fig:rmse}%
\end{figure}

The advantages of the centroid virial estimators compared with thermodynamic
estimators are shown in Fig.~\ref{fig:rmse}. Remarkably, the statistical
errors behave similarly in the PA and TI scheme (except for the lowest $P$
values). In both factorizations, the statistical error of the thermodynamic
estimator grows with $P$, whereas the error of the centroid virial estimator
is approximately independent of $P$. This means that if the centroid virial
estimator is used, the number of Monte Carlo samples need not be increased (at
least not significantly) when $P$ is increased in order to reach the quantum
limit. The overall conclusion of Figs.~\ref{fig:dFdl}-\ref{fig:rmse} is that
the optimal PIMC approach is using the TI\ factorization together with the
centroid virial estimator: the former lowers the discretization error while
the latter decreases the statistical error.

\begin{figure}[ptbh]
\includegraphics[width=\hsize]{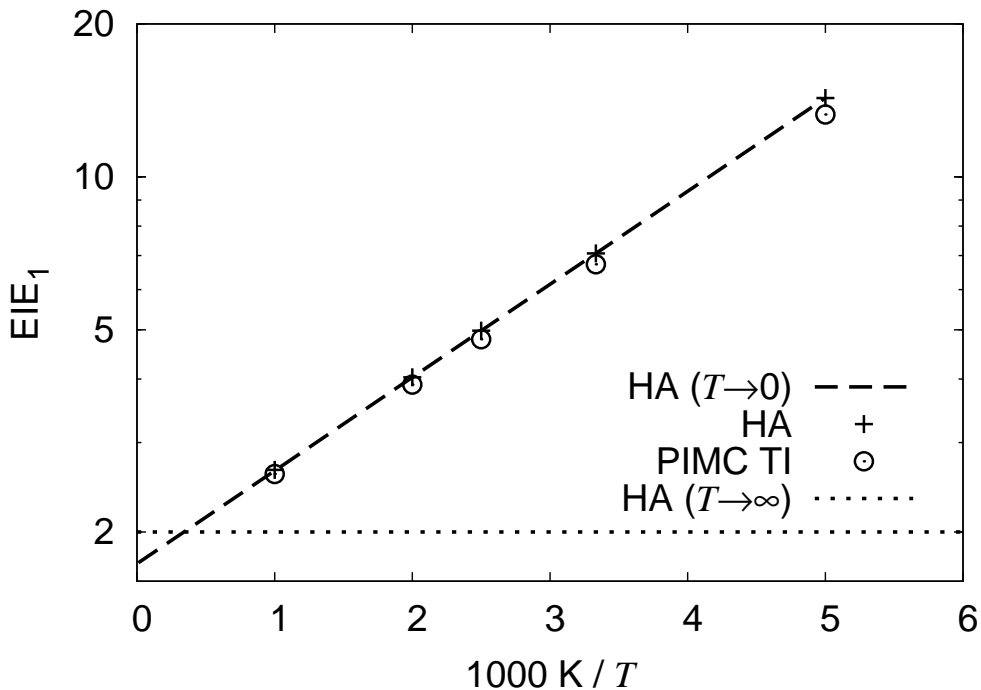}
\caption{Temperature dependence of the equilibrium isotope effect EIE$_{1}%
$~(\ref{eie1}). Comparison between the harmonic approximation (HA) and the
present PIMC method using the Takahashi-Imada factorization (PIMC TI).}%
\label{fig:eie}%
\end{figure}

Values of EIE$_{1}$ at different temperatures with their statistical errors
(computed by block-averaging \cite{Flyvbjerg_Petersen:1989}) are given in
Table~\ref{table:eie1} and plotted in Fig.~\ref{fig:eie}. Path integral
results are compared with the harmonic approximation (\ref{ha}) and its high
and low temperature limits [Eqs.~(\ref{ha_highT}) and (\ref{ha_lowT})]. The
table demonstrates that both the PA and TI factorizations, and both the
thermodynamic and centroid virial estimators converge to the same result,
which differs from the harmonic approximation. The only, yet crucial
differences, among the four PIMC results are in the required value of $P$
(which is lower in the TI\ scheme than in the PA) and in the statistical error
(which is much lower for the centroid virial than for the thermodynamic
estimators). \textbf{Surprisingly, the PI correction to the harmonic
approximation of the EIE is rather small; this was observed before, e.g., in
the EIE on three sigmatropic hydrogen shift reactions
\cite{Zimmermann_Vanicek:2009}. The reason is that the quantum harmonic
approximation works rather well in rigid molecules. The PI corrections would
be more pronounced in the EIE in floppy and other very anharmonic systems, or
in the kinetic isotope effect on reactions with significant tunneling
contribution.}

\begin{table}[ptbh]
\caption{Equilibrium isotope effect EIE$_{1}$~(\ref{eie1}) at several
temperatures. Table compares results of PIMC calculations using either the
primitive approximation (PA) or Takahashi-Imada (TI) factorizations, and
either the thermodynamic (TE) or centroid-virial (CVE) estimators. $P$ is the
Trotter number and the statistical error is shown in parentheses. Results of
the direct PIMC\ approach (\ref{eie1-direct}) and of the harmonic
approximation (HA) are shown as well.}
\begin{tabular}
[c]{ccccccccc}\hline\hline
$T(\text{K})$ & \multicolumn{3}{c}{PIMC PA} & \multicolumn{4}{c}{PIMC TI} &
\multicolumn{1}{c}{\text{HA}}\\\cline{2-4}\cline{5-8}
& \multicolumn{1}{c}{\text{TE}} & \multicolumn{1}{c}{\text{CVE}} & $P$ &
\multicolumn{1}{c}{\text{TE}} & \multicolumn{1}{c}{\text{CVE}} &
\multicolumn{1}{c}{\text{CVE direct}} & $P$ & \\\hline
$200$ & 13.1(3) & 13.24(3) & $160$ & 13.3(2) & 13.27(3) & 13.19(4) & $48$ &
14.29\\
$300$ & 6.82(8) & 6.72(1) & $110$ & 6.66(5) & 6.73(1) & 6.70(1) & $32$ &
7.07\\
$400$ & 4.84(4) & 4.778(2) & $80$ & 4.78(2) & 4.788(3) & 4.770(3) & $24$ &
4.98\\
$500$ & 3.91(2) & 3.896(1) & $60$ & 3.91(1) & 3.905(2) & 3.892(2) & $18$ &
4.03\\
$1000$ & 2.605(9) & 2.6018(3) & $30$ & 2.600(7) & 2.6036(4) & 2.5966(4) & $8$
& 2.65\\\hline\hline
\end{tabular}
\label{table:eie1}%
\end{table}

These differences are reflected in the simulation lengths.
Table~\ref{table:speedup1} shows speedups achieved using different methods in
comparison with the simulation using the PA and the thermodynamic estimator.
The speedup $f$ is defined as%
\begin{equation}
f_{\text{method}}:=\left(  \frac{\sigma_{\text{PA+TE}}}{\sigma_{\text{method}%
}}\right)  ^{2}\frac{t_{\text{PA+TE}}}{t_{\text{method}}},\label{eq:speedup}%
\end{equation}
where \textquotedblleft method\textquotedblright\ stands for the factorization
and estimator used (TE = thermodynamic estimator), $t$ is the CPU time taken
by the simulation, and $\sigma$ is the statistical error achieved. The number
of beads, $P$, was chosen so that the discretization error was roughly the
same for the PA and TI factorization. The thermodynamic and centroid virial
estimators were computed together in a single simulation, which is the reason
why the factor involving statistical errors appears in Eq.~(\ref{eq:speedup}).
Table~\ref{table:speedup1} shows that TI factorization alone accelerates
simulations by a factor between 5 and 10. If the TI factorization is augmented
with the centroid virial estimator, the overall acceleration is 200--2000
fold. \textbf{Surprisingly, the speedups due to the use of the centroid virial
estimator at low temperatures are smaller than at high temperatures; this
could be because at high temperatures the system accesses flatter regions of
the PES and because the centroid virial estimator yields (in the PA) the exact
result for the free particle without any sampling.}


\begin{table}[ptbh]
\caption{Speedup achieved in computing EIE$_{1}$~(\ref{eie1}) by various
methods in comparison with the time taken by the primitive approximation (PA)
with the thermodynamic estimator (TE). E.g., at $1000$ K, the calculation with
the Takahashi-Imada (TI)\ factorization and the centroid-virial estimator
(CVE) is 2125 times faster.}%
\label{table:speedup1}
\begin{tabular}
[c]{ccccc}\hline\hline
$T(\text{K})$ & \multicolumn{2}{c}{PA} & \multicolumn{2}{c}{TI}\\
& TE & CVE & TE & CVE\\\hline
$200$ & $1$ & $148$ & $9$ & $285$\\
$300$ & $1$ & $159$ & $7$ & $476$\\
$400$ & $1$ & $347$ & $9$ & $582$\\
$500$ & $1$ & $370$ & $10$ & $403$\\
$1000$ & $1$ & $865$ & $5$ & $2125$\\\hline\hline
\end{tabular}
\end{table}


Results for EIE$_{2}$ are displayed in Table~\ref{table:eie2}. Everything said
about EIE$_{1}$ holds here, \textbf{and in addition} the PIMC results also
agree with the harmonic approximation. The reason is probably the cancellation
of errors due to the harmonic approximation between the ratios $Q_{\text{D}%
_{2}}/Q_{\text{HD}}$ and $Q_{\text{HD}}/Q_{\text{H}_{2}}$.

\textbf{The results of EIE}$_{2}$\textbf{ were also found in a reasonable
agreement with the experimental data of Urey \textit{et al.} \cite{Urey:1934}.
Those values are shown in Table \ref{table:eie2} (in the column EXP.) and were
obtained by linear regression of the dependence of EIE}$_{2}$\textbf{ from
Ref. \cite{Urey:1934} as a function of }$1/T$\textbf{\ between the
temperatures of $25^{\circ}$C and $468^{\circ}$C; the datapoint at the
intermediate temperature from Ref. \cite{Urey:1934} was omitted as too
uncertain and the average of the two available results at $468^{\circ}$C was
used.}

The speedups achieved with the TI factorization and centroid virial estimator
are shown in Table~\ref{table:speedup2} and have the same order of magnitude
as those for EIE$_{1}$ in Table~\ref{table:speedup1}.

\begin{table}[ptbh]
\caption{Equilibrium isotope effect EIE$_{2}$~(\ref{eie2}) at several
temperatures. See caption of Table \ref{table:eie1} for details.
\textbf{Column EXP. contains values obtained by linear regression of
experimental results \cite{Urey:1934}.}}
\begin{tabular}
[c]{ccccccccc}\hline\hline
$T(\text{K})$ & \multicolumn{3}{c}{PIMC PA} & \multicolumn{3}{c}{PIMC TI} &
\multicolumn{1}{c}{\text{HA}} & \multicolumn{1}{c}{\text{EXP.}}\\\cline{2-4}%
\cline{5-7}
& \multicolumn{1}{c}{\text{TE}} & \multicolumn{1}{c}{\text{CVE}} & $P$ &
\multicolumn{1}{c}{\text{TE}} & \multicolumn{1}{c}{\text{CVE}} & $P$ &  &
\\\hline
$200$ & 2.83(6) & 2.85(1) & $160$ & 2.85(3) & 2.84(1) & $48$ & 2.86 & 2.89\\
$300$ & 3.31(3) & 3.256(4) & $110$ & 3.24(2) & 3.251(4) & $32$ & 3.26 & 3.28\\
$400$ & 3.54(3) & 3.473(2) & $80$ & 3.49(1) & 3.474(3) & $24$ & 3.49 & 3.48\\
$500$ & 3.63(2) & 3.618(2) & $60$ & 3.60(1) & 3.616(2) & $18$ & 3.63 & 3.60\\
$1000$ & 3.89(1) & 3.897(1) & $30$ & 3.90(1) & 3.894(1) & $8$ & 3.90 &
3.83\\\hline\hline
\end{tabular}
\label{table:eie2}%
\end{table}


\begin{table}[ptbh]
\caption{Speedup achieved in computing EIE$_{2}$~(\ref{eie2}) by various
methods. See caption of Table \ref{table:speedup1} for details.}%
\label{table:speedup2}
\begin{tabular}
[c]{ccccc}\hline\hline
$T(\text{K})$ & \multicolumn{2}{c}{PA} & \multicolumn{2}{c}{TI}\\
& TE & CVE & TE & CVE\\\hline
$200$ & $1$ & $54$ & $10$ & $146$\\
$300$ & $1$ & $63$ & $10$ & $229$\\
$400$ & $1$ & $150$ & $9$ & $233$\\
$500$ & $1$ & $154$ & $9$ & $217$\\
$1000$ & $1$ & $243$ & $5$ & $1012$\\\hline\hline
\end{tabular}
\end{table}


As mentioned above, both EIE$_{1}$ and EIE$_{2}$ are equilibrium constants of
isotopomerization reactions, hence the present method can be viewed as
complementary to the PIMC method for calculating the temperature dependence of
the equilibrium constant described in Ref.~\cite{Buchowiecki:2012}. In
Ref.~\cite{Buchowiecki:2012}, the equilibrium constant $K(T)$ at temperature
$T$ is obtained by thermodynamic integration with respect to the inverse
temperature from the equilibrium constant $K(T_{0})$ at temperature $T_{0}$.
In isotopomerization reactions, $K(T_{0}\rightarrow\infty)$ at high
temperatures is simply given by the ratio of the symmetry numbers, which
follows easily from Eq. (\ref{ha_highT}).

Note also that the second alternative form of the EIE$_{2}$ in Eq.~(\ref{eie2}%
) involves separate reactant or product molecules, and can be used with
molecular potential energy surfaces if the full reactive potential energy
surface is not available.


\section{\label{sec:conclusions}Conclusions}

Using two isotopomerization reactions, we have demonstrated that TI
factorization significantly decreases the Trotter number required for
convergence of PIMC calculations of EIEs to the quantum limit. This leads to
significant acceleration of the calculations whether forces are available in
an analytical form, or must be computed by finite differences from energies.
Similar acceleration was observed in
Refs.~\cite{Azuri:2011,Perez_Tuckerman:2011}. Moreover, we have observed that
the large difference in the statistical convergence of the thermodynamic and
centroid virial estimators, well known from the PIMC calculations based on the
PA, reappears in the TI\ scheme. Unlike the thermodynamic estimator, the
centroid virial estimator has a statistical error independent of the Trotter
number. As shown in Sec.~\ref{sec:results}, this can easily accelerate
calculations by orders of magnitude.

The TI scheme applies, unfortunately, only to the diagonal matrix elements of
the density matrix. If the off-diagonal elements are needed, as in computing
the momentum distribution functions, alternative high order factorizations are
required \cite{Perez_Tuckerman:2011}. \textbf{An example is the Suzuki-Chin
factorization \cite{Suzuki:1995,Chin:1997}. Indeed, one of us is presently
exploring how this factorization could be used to efficiently compute the
kinetic isotope effects as opposed to the EIE.} In calculations of diagonal
elements, several Chin factorizations \cite{Chin:2004} may converge faster
than the TI splitting although they have the same order in general
\cite{Azuri:2011}. Yet, the TI scheme has one decisive advantage, which is its
simplicity. As demonstrated here, the TI scheme is easily implemented in
existing codes based on the PA: the implementation simply replaces the
original potential $V$ by the TI effective potential $V_{\text{eff}}.$ Unlike
some more sophisticated factorizations, the TI scheme treats all imaginary
time slices equally. In conclusion, we believe that the TI factorization will
find more interesting applications in path integral\ simulations of
equilibrium and nonequilibrium quantum effects.

\section{Acknowledgments}

This research was supported by the Swiss NSF Grant No. $200021\_124936/1$ to
J.V. and by the EPFL. We thank Matthew Wodrich for a careful reading of the manuscript.



\end{document}